\documentclass[final,authoryear,5p,twocolumn,times]{elsarticle}
\usepackage{graphicx,subfigure}
\usepackage{color}
\usepackage{rotating}
\usepackage{subfigure}
\usepackage{amssymb}
\usepackage{amsthm,amsmath}

\newtheorem{thm}{Theorem}

\newtheorem{rem}{Remark}

\newcommand{\ben}{\begin{enumerate}}
\newcommand{\een}{\end{enumerate}}
\newcommand{\bit}{\begin{itemize}}
\newcommand{\eit}{\end{itemize}}
\newcommand{\be}{\begin{equation}}
\newcommand{\ee}{\end{equation}}
\newcommand{\bdm}{\begin{displaymath}}
\newcommand{\edm}{\end{displaymath}}
\newcommand{\bea}{\begin{eqnarray}}
\newcommand{\eea}{\end{eqnarray}}

\begin{document}

%

\begin{frontmatter}

\title{Molecular Distributions in Gene Regulatory Dynamics}

\author{Michael C. Mackey}

\address{Departments of Physiology, Physics \& Mathematics and Centre for
Nonlinear Dynamics, McGill University, 3655 Promenade Sir William Osler, Montreal, QC, CANADA, H3G 1Y6}

\ead{mackey@cnd.mcgill.ca}

\author{Marta Tyran-Kami\'nska}

\address{Institute of Mathematics,
University of Silesia, Bankowa 14, 40-007 Katowice, POLAND}

\ead{mtyran@us.edu.pl}

\author{Romain Yvinec\corref{cor1}}

\address{Universit\'{e} de Lyon
      CNRS
             Universit\'{e} Lyon 1 ,
              B\^{a}t Braconnier
        43 bd du 11 nov. 1918
        F-69622 Villeurbanne Cedex
        France}

\cortext[cor1]{Corresponding author}

\ead{yvinec@math.univ-lyon1.fr}

\begin{abstract}  We show how one may analytically compute the stationary density of the distribution of molecular constituents in
populations of cells in the presence of noise arising from either bursting transcription or translation, or noise
in degradation rates arising from low numbers of molecules.  We have compared our results with an analysis of the same model systems (either inducible or repressible operons) in the absence of any stochastic effects, and shown the correspondence between behaviour in the deterministic system and the stochastic analogs.  We have identified key dimensionless parameters that control the appearance of one or two steady states in the deterministic case, or unimodal and bimodal densities in the stochastic systems, and detailed the analytic requirements for the occurrence of different behaviours.  This approach provides, in some situations,
an alternative to computationally intensive stochastic simulations.  Our results indicate that,
within the context of the simple models we have examined, bursting and degradation noise cannot be distinguished
analytically when present alone.
\end{abstract}

\begin{keyword}
Stochastic modelling, inducible/repressible operon.
\end{keyword}

\end{frontmatter}


\section{Introduction}\label{sec:intro}

In neurobiology, when it became clear that some of the fluctuations seen in whole nerve recording, and later in single cell recordings, were not simply measurement noise but actual fluctuations in the system being studied, researchers very quickly started wondering to what extent these fluctuations actually played a role in the operation of the nervous system.

Much the same pattern of development has occurred in cellular and molecular biology as experimental techniques have allowed investigators to probe temporal behaviour at ever finer levels, even to the level of individual molecules.  Experimentalists and theoreticians alike who are interested in the regulation of gene networks are increasingly focussed on trying to access the role of various types of fluctuations on the operation and fidelity of both simple and complex gene regulatory systems.  Recent reviews \citep{kaern05,raj08,swain08c} give an interesting perspective on some of the issues confronting both experimentalists and modelers.

Typically, the discussion seems to focus on whether fluctuations  can be considered as extrinsic to the system
under consideration \citep{swain08a,ochab08,ochab10}, or whether they are an intrinsic part of the fundamental
processes they are affecting (e.g. bursting, see below). The dichotomy is rarely so sharp however, but
\citet{elowitz02} have used an elegant experimental technique to distinguish between the two, see also
\citet{raser04}, while \citet{swain02a} and \citet{scott06} have laid the groundwork for a theoretical
consideration of this question.  One issue that is raised persistently in considerations of the role of
fluctuations or noise in the operation of gene regulatory networks is whether or not they are ``beneficial"
\citep{blake06} or ``detrimental" \citep{fraser04} to the operation of the system under consideration.  This is,
of course, a question of definition and not one that we will be further concerned with here.

Here, we consider in detail the density of the molecular distributions  in generic bacterial operons in the
presence of `bursting' (commonly known as intrinsic noise in the biological literature) as well as inherent
(extrinsic) noise using an analytical approach. Our work is motivated by the well documented production of mRNA
and/or protein in stochastic bursts in both prokaryotes and eukaryotes
\citep{blake03,cai,chubb,golding,raj,sigal,yu}, and follows other contributions by, for example,
\cite{kepler01}, \cite{friedman06}, \cite{rudnicki07} and \cite{swain08b}.


In  Section \ref{sec:generic} we develop the concept of the operon and treat simple models of the classic
inducible and repressible operon. Section \ref{sec:dynamics-single} considers the effects of bursting alone in
an ensemble of single cells.  Section \ref{sec:dynamics-degrad} then examines the situation in which there are
continuous white noise fluctuations in the dominant
species degradation rate in the absence of  bursting. 

\section{Generic operons}\label{sec:generic}

\subsection{The operon concept}\label{ssec:concept}

The so-called `central dogma' of molecular biology is simple to
state in principle, but complicated in its detail.  Namely through
the process of {\it transcription} of DNA,  messenger RNA (mRNA,
$M$) is produced and, in turn, through the process of {\it
translation} of the mRNA proteins (intermediates, $I$) are
produced. There is often feedback in the sense that molecules
(enzymes, $E$) whose production is controlled by these proteins
can modulate the translation and/or transcription processes.  In
what follows we will refer to these molecules  as {\it effectors}.
We now consider both the transcription and translation process in
more detail.

In the transcription process an amino acid sequence in the DNA is
copied by the enzyme RNA polymerase (RNAP) to produce a
complementary copy of the DNA segment encoded in the resulting
RNA.  Thus this is the first step in the transfer of the
information encoded in the DNA.  The process by which this occurs
is as follows.

When the DNA is in a double stranded configuration, the RNAP is
able to recognize and bind to the promoter region of the DNA. (The
RNAP/double stranded DNA complex is known as the closed complex.)
Through the action of the RNAP, the DNA is unwound in the vicinity
of the RNAP/DNA promoter site, and becomes single stranded.  (The
RNAP/single stranded DNA is called the open complex.)  Once in the
single stranded configuration, the transcription of the DNA into
mRNA commences.

In prokaryotes, translation of the newly formed mRNA commences
with the binding of a ribosome to the mRNA.  The function of the
ribosome is to `read' the mRNA in triplets of nucleotide sequences
(codons).  Then through a complex sequence of events,  initiation
and elongation factors bring transfer RNA (tRNA) into contact with
the ribosome-mRNA complex to match the codon in the mRNA to the
anti-codon in the tRNA.  The elongating peptide chain consists of
these linked amino acids, and it starts folding into its final
conformation.  This folding continues until the process is
complete and the polypeptide chain that results is the mature
protein.

The lactose ({\it lac}) operon in bacteria is the paradigmatic
example of this concept and this much studied system consists of
three structural genes named {\it lacZ}, {\it lacY}, and {\it
lacA}.  These three genes contain the code for the ultimate
production, through the translation of mRNA,  of the intermediates
$\beta$-galactosidase, {\it lac} permease, and thiogalactoside
transacetylase respectively.  The enzyme $\beta$-galactosidase is
active in the conversion of lactose into allolactose and then the
conversion of allolactose into glucose. The {\it lac} permease is
a membrane protein responsible for the transport of extracellular
lactose to the interior of the cell. (Only the transacetylase
plays no apparent role in the regulation of this system.) The regulatory
gene {\it lacI}, which is part of a different operon, codes for
the {\it lac} repressor, which is transformed to an inactive form
when bound with allolactose, so in this system allolactose
functions as the effector molecule.

\subsection{The transcription rate function}
\label{ssec:response}

In this section we examine the molecular dynamics of both the
classical inducible and repressible operon to derive expressions
for the dependence of  the transcription  rate on effector levels.
(When  the transcription rate is constant and independent of the
effector levels we will refer to this as the no control
situation.)

\subsubsection{Inducible regulation} For a typical inducible regulatory situation (such as the {\it lac} operon),
in the {\it presence} of the effector
molecule the repressor is {\it inactive} (is unable to bind to the
operator region preceding the structural genes), and thus DNA
transcription can proceed. Let $R$ denote the repressor, $E$ the
effector molecule,  and $O$ the operator. The effector is known to
bind with the active form $R$ of the repressor. We assume  that
this reaction is of the form
\begin{equation}
R + nE \stackrel{K_1} \rightleftharpoons RE_n \qquad K_1 = \frac{RE_n}{R \cdot E^n} \label{ind-re},
\end{equation}
where $n$ is the effective number  of molecules of effector
required to inactivate the repressor $R$.  Furthermore, the
operator $O$ and repressor $R$ are assumed to interact according
to
\begin{equation*}
O + R \stackrel{K_2}\rightleftharpoons OR \qquad K_2 = \frac{OR}{O \cdot R}. \label{ind-or}
\end{equation*}
Let the total operator be $O_{tot}$:
\begin{equation*}
O_{tot} = O + OR = O + K_2 O \cdot R = O(1+K_2R), \label{ind-totoper}
\end{equation*}
and the total level of repressor be $R_{tot}$:
\begin{equation*}
R_{tot} = R + K_1 R \cdot E^n + K_2 O \cdot R. \label{ind-totrepre}
\end{equation*}
The fraction of operators not bound by repressor (and therefore
free to synthesize mRNA) is given by
\begin{equation*}
f(E) = \frac{O}{O_{tot}} = \frac{1}{1+K_2R}. \label{ind-frac1}
\end{equation*}
If the amount  of repressor $R$ bound to the operator $O$ is small
\begin{equation*}
R_{tot} \simeq R + K_1 R \cdot E^n = R(1+K_1E^n)
\label{ind-rtotapprox}
\end{equation*}
so
\begin{equation*}
R = \frac {R_{tot}}{1+K_1E^n}, \label{ind-R}
\end{equation*}
and consequently
\begin{equation}
f(E) = \frac{1+ K_1E^n}{1+K_2R_{tot} + K_1E^n} = \frac{1+
K_1E^n}{K + K_1E^n}, \label{ind-frac2}
\end{equation}
where $K = 1+K_2R_{tot}$.  There will be maximal repression when
$E=0$ but even then  there will still be a basal level of mRNA
production  proportional to $K^{-1}$ (which we call the fractional
leakage).

If the maximal DNA transcription rate is $\bar \varphi_{m}$ (in
units of inverse time) then, under the assumption that the rate of
transcription $\varphi$ in the entire population is proportional
to the fraction $f$ of unbound operators,  the variation $\varphi$
of the DNA transcription rate with the effector level is given by
$\varphi = \bar \varphi_{m} f$, or
    \be
    \varphi(E) = \bar \varphi_{m}    \frac{1+
    K_1E^n}{K + K_1E^n}.
     \label{ind-phi}
\ee

\subsubsection{Repressible regulation} In the classic example of a repressible system
(such as the {\it trp} operon) in the {\it presence} of the
effector molecule the repressor is {\it active} (able to bind to
the operator region), and thus block DNA transcription. We use the
same notation as before, but now note that the effector binds with
the inactive form $R$ of the repressor so it becomes active. We
assume that this reaction is of the same form as in Equation
\ref{ind-re}.  The difference now is that the operator $O$ and
repressor $R$ are assumed to interact according to
\begin{equation*}
O + R\cdot E^n \stackrel{K_2}\rightleftharpoons ORE_n \qquad K_2 = \frac{ORE_n}{O
\cdot R\cdot E_n}. \label{or-rep}
\end{equation*}
The  total operator is now given by
\begin{equation*}
O_{tot} = O + ORE_n = O + K_2 O \cdot R\cdot E^n = O(1+K_2R\cdot E^n), \label{rep-totoper}
\end{equation*}
so the fraction of operators not bound by repressor is given by
\begin{equation*}
f(E) = \frac{O}{O_{tot}} = \frac{1}{1+K_2R\cdot E^n}. \label{rep-frac1}
\end{equation*}
Again assuming that the amount  of repressor $R$ bound to the
operator $O$ is small we have
\begin{equation*}
f(E) = \frac{1+ K_1E^n}{1+(K_1+K_2R_{tot})E^n} = \frac{1+
K_1E^n}{1 + KE^n}, \label{rep-frac2}
\end{equation*}
where  $K = K_1+K_2R_{tot}$. Now  there will be maximal repression when $E$ is large, but even at maximal
repression there will still be a basal level of mRNA production proportional to $K_1 K^{-1}<1$. The variation of
the DNA transcription rate with effector level is given by $\varphi = \bar \varphi_{m} f$ or
    \be
    \varphi(E) = \bar \varphi_{m}  \frac{1+
    K_1E^n}{1 + KE^n}.
    \label{rep-phi}
    \ee

Both (\ref{ind-phi}) and (\ref{rep-phi})  are special cases of the
function
    \be
    \varphi(E) = \bar \varphi_{m} \dfrac{1+K_1E^n}{A+BE^n} =
    \bar \varphi_{m}f(E).
      \label{eq:gen-response-fun}
    \ee
where $A,B  \geq 0$ are given in Table \ref{tab:ABC1}.

\begin{table}
\centering
\begin{tabular}{|c|c|c|}
  \hline
  \rule[-1ex]{0.0cm}{4ex} parameter  & inducible & repressible \\
  \hline   \hline
\rule[-1ex]{0.0cm}{4ex}  $A$ & $K=1+K_2R_{tot}$ & $1$  \\ \hline \rule[-1ex]{0.0cm}{4ex}  $B $& $K_1$ &
$K=K_1+K_2R_{tot} $ \\ \hline \rule[-3ex]{0.0cm}{7ex}  $\dfrac BA$ & $\dfrac{K_1}{K}$ & $K$ \\ \hline
\rule[-3ex]{0.0cm}{7ex}    $\Lambda = A$ & $K $ & $1$ \\ \hline
\rule[-3ex]{0.0cm}{7ex} $\Delta = B K_1^{-1} $ & $1$ & $KK_1^{-1}$ \\
\hline \rule[-3ex]{0.0cm}{7ex}  $\theta = \dfrac{\kappa_d}{n\Delta}\left(1 - \dfrac \Delta \Lambda  \right )$ &
$\dfrac{\kappa_d }{n}\cdot \dfrac {K-1}{K} > 0$
  & $\dfrac{\kappa_d }{n}\cdot \dfrac {K_1-K}{K} < 0$ \\
  \hline
\end{tabular}
\caption{Definitions of the parameters $A$, $B$, $\Lambda$, $\Delta$ and $\theta$.   See the text and Section
\ref{ssec:response} for more detail.} \label{tab:ABC1}
\end{table}

\subsection{Deterministic operon dynamics in a population of cells}\label{ssec:dynamics-ensemble}

The reader may wish to consult \citet{polynikis09} for an interesting survey of techniques applicable to this approach.

We first consider a large population of cells, each of which
contains one copy of a particular operon, and let $(M,I,E)$ denote
mRNA, intermediate protein, and effector levels respectively {\it
in the population}. Then for a generic operon with a maximal level
of transcription $ \bar b_{d}$ (in concentration units), we have
dynamics described by the system
\citep{Griffith68a,Griffith68b,othmer76,selgrade79}
   \begin{align}
    \dfrac{dM}{dt} &= \bar b_{d} \bar \varphi_{m}f( E) -\gamma_M M,\label{eq:mrna}\\
    \dfrac{dI}{dt} &= \beta_I    M -\gamma_I I ,\label{eq:intermed}\\
    \dfrac{dE}{dt} &= \beta_E   I -\gamma_E E.\label{eq:effector}
    \end{align}
Here we assume that the rate of mRNA production is proportional to the fraction of time the operator region is
active, and that the rates of intermediate and enzyme production are simply proportional to the amount of mRNA
and intermediate respectively. All three of the components $(M,I,E)$ are subject to random loss. The function $f$ is
calculated in the previous section.

It will greatly simplify matters to rewrite Equations
\ref{eq:mrna}-\ref{eq:effector} by defining dimensionless
concentrations.  To this end we first rewrite Equation
\ref{eq:gen-response-fun} in the form
    \be
    \varphi(e) =
    \varphi_{m}f(e),
    \label{eq:gen-response-fun-dimen}
    \ee
where $\varphi_m$ (dimensionless) is defined by
    \be
    \varphi_m = \dfrac {\bar \varphi_m \beta_E \beta_I}{\gamma_M \gamma_E \gamma_I} \quad \mbox {and} \quad f(e)=
    \dfrac{1+e^n}{\Lambda + \Delta e^n},
    \label{eq:gen-response-fun-dimen-f}
    \ee
$\Lambda$ and $\Delta$ are defined in Table \ref{tab:ABC1},
and we have defined a dimensionless effector concentration $(e)$
through
    \begin{equation*}
    E = \eta e   \quad \mbox{with} \quad\eta = \dfrac {1}{\sqrt[n]{K_1}}.
    \end{equation*}
Further defining dimensionless intermediate  ($i$) and mRNA
concentrations ($m$) through
    \begin{equation*}
    I = i\eta \dfrac{\gamma_E}{\beta_E} \quad\text{and}\quad M = m
    \eta \dfrac{\gamma_E \gamma_I}{\beta_E \beta _I},
    \end{equation*}
Equations \ref{eq:mrna}-\ref{eq:effector} can be written in the
equivalent form
 \begin{align*}
    \dfrac{dm}{dt} &= \gamma_M [ \kappa_d f( e) - m],\\
    \dfrac{di}{dt} &= \gamma_I ( m - i),  \\
    \dfrac{de}{dt} &= \gamma_E (i - e),
    \end{align*}
where
    \be
    \kappa_d = b_d \varphi_m \quad \mbox{and} \quad b_d =\dfrac {\bar b_d}{\eta}
    \label{eq:kappa}
    \ee
are dimensionless constants.

For notational simplicity, henceforth we denote dimensionless
concentrations by $(m,i,e) = (x_1,x_2,x_3)$, and subscripts
$(M,I,E) = (1,2,3)$.  Thus we have
    \begin{align}
    \dfrac{dx_1}{dt} &= \gamma_1 [\kappa_d f( x_3) -x_1], \label{eq:mrna1-dimen}\\
    \dfrac{dx_2}{dt} &= \gamma_2  (  x_1 - x_2), \label{eq:intermed2-dimen}\\
    \dfrac{dx_3}{dt} &= \gamma_3  ( x_2 - x_3).\label{eq:effector3-dimen}
    \end{align}
In each equation, $\gamma_i  $ for $i=1,2,3$ denotes a net loss
rate (units of inverse time), and thus   Equations
\ref{eq:mrna1-dimen}-\ref{eq:effector3-dimen} are not in
dimensionless form.

The dynamics of this classic operon model can be fully analyzed. Let $X=(x_1,x_2,x_3)$ and denote by $S_t(X)$
the flow generated by the system (\ref{eq:mrna1-dimen})-(\ref{eq:effector3-dimen}).  For both inducible and
repressible operons, for all initial conditions $X^0=(x_1^0,x_2^0,x_3^0) \in \mathbb{R}_3^+$ the flow $S_t(X^0)
\in \mathbb{R}_3^+$ for   $t>0$.

Steady states of the system (\ref{eq:mrna1-dimen})-(\ref{eq:effector3-dimen}) are in a one to one correspondence
with solutions of the equation
    \be
    \dfrac {x}{\kappa_d} = f(x)
    \label{eq:ss-dimen}
    \ee
and for each solution $x^*$ of Equation \ref{eq:ss-dimen} there is
a steady state $X^*= (x_1^*,x_2^*,x_3^*)$ of
(\ref{eq:mrna1-dimen})-(\ref{eq:effector3-dimen}) given by
    \begin{equation*}
    x_1^* = x_2^* = x_3^* = x^*.
    \label{eq:ss-full-dimen}
    \end{equation*}
Whether there is a single steady state $X^*$ or there are multiple
steady states will depend on whether we are considering a
repressible or inducible operon.

\subsubsection{No control}  In this case, $f(x) \equiv 1$, and there is
a single steady state $x^* = \kappa_d$ that is globally
asymptotically stable.

\subsubsection{Inducible regulation}\label{sss:induc-deter}

\begin{figure}
\centering
\includegraphics[width=\columnwidth]{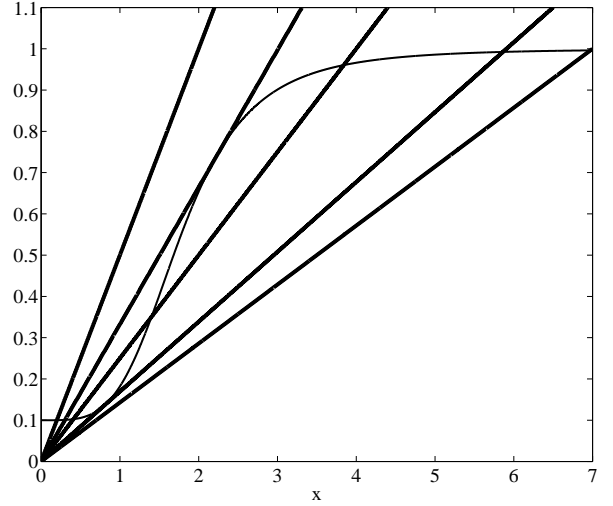}

\caption{Schematic illustration of the possibility of one, two or
three solutions of Equation \ref{eq:ss-dimen} for varying
values of $\kappa_d$ with inducible regulation.   The
monotone increasing graph is the function $f$ of Equation \ref{eq:gen-response-fun-dimen-f}, and the
straight lines correspond to $x/\kappa_d$ for (in a clockwise direction) $\kappa_d \in [0,\kappa_{d-})$,
$\kappa_d = \kappa_{d-}$,
$\kappa_d \in (\kappa_{d-},\kappa_{d+})$,
$\kappa_d = \kappa_{d+}$,
and $\kappa_{d+} < \kappa_d$.
This figure was constructed with $n=4$ and $K=10$ for which $\kappa_{d-}=3.01$ and
$\kappa_{d+} = 5.91$ as computed from (\ref{eq:kappa-pm}).  See the text for further details.  } \label{fig:123SS}
\end{figure}

\begin{figure}[t]
\centering
\includegraphics[width=\columnwidth]{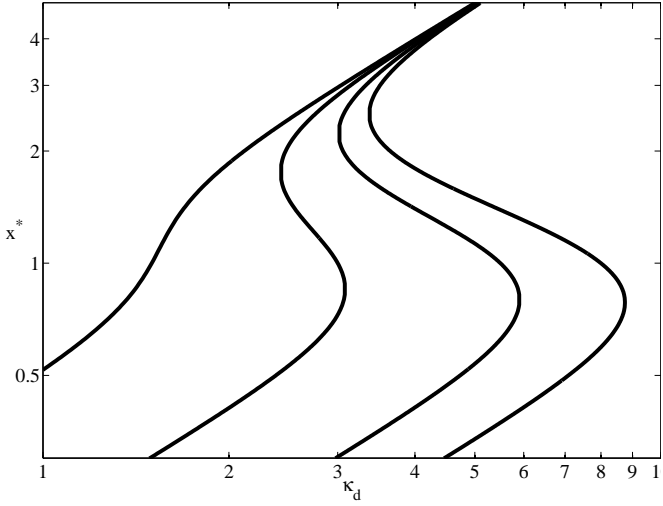}
\caption{Full logarithmic plot of the steady state values of $x^*$ versus
$\kappa_d$ for an inducible system, obtained from Equation \ref{eq:ss-dimen},
for $n=4$ and $K = 2,5,10,\mbox{and}\,\, 15$   (left to right) illustrating the
dependence of the occurrence of bistability on $K$. See the text for details.}
\label{fig:deter-inducible-1}
\end{figure}

\noindent{\bf Single versus multiple steady states.} For an
inducible operon with $f$ given by Equation \ref{ind-frac2}, there
may be one ($X_1^*$ or $X_3^*$), two ($X_1^*,X_2^*=X_3^*$ or
$X_1^*=X_2^*,X_3^*$), or three ($X_1^*,X_2^*,X_3^*$) steady
states, with the ordering $0< X_1^* \leq X_2^* \leq X_3^*$,
corresponding to the possible solutions of Equation
\ref{eq:ss-dimen} (cf. Figure \ref{fig:123SS}).  The smaller
steady state $(X_1^*)$ is typically referred to as an uninduced
state, while the largest steady state $(X_3^*)$ is called the induced
state. The steady state values of $x$ are easily obtained from
(\ref{eq:ss-dimen}) for given parameter values, and the dependence
on $\kappa_d$ for $n=4$ and a variety of values of $K$ is shown in
Figure \ref{fig:123SS}. Figure \ref{fig:deter-inducible-1} shows a
graph of the steady states $x^*$ versus $\kappa_d$ for various
values of the leakage parameter $K$.

Analytic conditions for the existence of one or more steady states
can be obtained by using Equation \ref{eq:ss-dimen} in conjunction
with the observation that the delineation points are marked by the
values of $\kappa_d$ at which $x/\kappa_d$ is tangent to $f(x)$
(see Figure \ref{fig:123SS}). Simple differentiation of
(\ref{eq:ss-dimen}) yields the second condition
    \be
    \dfrac 1 {\kappa_d n(K-1 ) }
    = \dfrac{ x^{n-1}}{(K+x^n)^2}.
    \label{eq:slope}
    \ee
From equations (\ref{eq:ss-dimen}) and (\ref{eq:slope}) we obtain
the values of $x$ at which tangency will occur:
    \be
    x_{\pm} = \sqrt[n]{\dfrac{K-1}{2 }
    \left\{
    \left[
    n - \dfrac {K+1}{K-1}
    \right]
    \pm \sqrt
    {
    n^2
    -2n\dfrac{K+1}{K-1}
    +1
 }
    \right\}}.
    \label{eq:tangency}
    \ee
The two corresponding values of $\kappa_d$ at which a tangency
occurs are given by
  \be
    \kappa_{d \pm} = x_{\mp} \dfrac{K+x_{\mp}^n}{1 + x_{\mp}^n}.
    \label{eq:kappa-pm}
        \ee
(Note the deliberate use of $x_{\mp}$ as opposed to $x_{\pm}$.)

\begin{figure}
\centering
\includegraphics[width=\columnwidth]{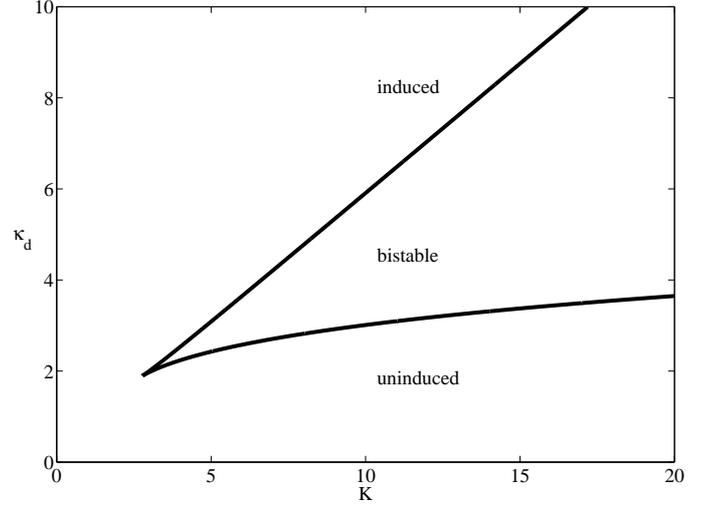}
\caption{In this figure we present a parametric plot (for $n=4$) of the bifurcation diagram in $(K,\kappa_d)$
parameter space delineating one from three steady states in a deterministic  inducible operon as obtained from
Equations \ref{eq:ss-dimen} and \ref{eq:slope}. The upper (lower) branch corresponds to $\kappa_{d-}$
($\kappa_{d+}$), and for all values of $(K,\kappa_d)$ in the interior of the cone  there are two locally stable
steady states $X_1^*,X_3^*$, while outside there is only one.  The tip of the cone occurs at
$(K,\kappa_d)=((5/3)^2,(5/3)\sqrt[4]{5/3})$ as given by Equations \ref{eq:K-necessary}  and
\ref{eq:kappa-necessary}. For $K\in [0,(5/3)^2)$ there is but a
single steady state. 
} \label{fig:contour-deter}
\end{figure}

A necessary condition for the existence of two or more steady
states is obtained by requiring that the square root in
(\ref{eq:tangency}) be non-negative, or
    \be
    K \geq \left ( \dfrac {n+1}{n-1}\right )^2.
    \label{eq:K-necessary}
    \ee
From this a second necessary condition follows, namely
    \be
    \kappa_d \geq \dfrac{n+1}{n-1}\sqrt[n]{\dfrac{n+1}{n-1}}.
    \label{eq:kappa-necessary}
    \ee
Further, from Equations \ref{eq:ss-dimen} and \ref{eq:slope} we
can delineate the boundaries in $(K,\kappa_d)$ space in which
there are one or three locally stable steady states as shown in
Figure \ref{fig:contour-deter}.  There, we have given a parametric
plot ($x$ is the parameter) of $\kappa_d$ versus $K$, using
    \begin{equation*}
    K(x)= \dfrac{x^n[x^n+(n+1)]}{(n-1)x^n-1} \quad \mbox{and}\quad
    \kappa_d(x)= \dfrac{[K(x)+x^n]^2}{nx^{n-1}[K(x)-1]},
    \label{eq:deter-induc-para}
    \end{equation*}
for $n=4$ obtained from Equations \ref{eq:ss-dimen} and \ref{eq:slope}. As is clear from the figure, when
leakage is appreciable (small $K$, e.g for $n=4$, $K < (5/3)^2$) then the possibility of bistable behaviour is
lost.

\begin{rem}  Some general observations on the influence of $n$,
$K$, and  $\kappa_d$ on the appearance of bistability in the
deterministic case are in order.
\begin{enumerate}
\item The degree of cooperativity $(n)$ in the binding of
    effector to the repressor plays a significant role.
    Indeed, $n>1$ is a necessary condition for bistability.
\item If $n>1$ then a second necessary condition for
    bistability is that $K$ satisfies Equation
    \ref{eq:K-necessary} so the fractional leakage $(K^{-1})$
    is sufficiently small.

\item Furthermore, $\kappa_d$ must satisfy Equation
    \ref{eq:kappa-necessary} which is quite instructive.
    Namely for $n \to \infty$ the limiting lower limit is
    $\kappa_d > 1$ while for $n \to 1$ the minimal value of
    $\kappa_d$ becomes quite large.  This simply tells us that
    the ratio of the product of the production rates to the
    product of the  degradation rates must always be greater
    than 1 for bistability to occur, and the lower the degree
    of cooperativity $(n)$ the larger the ratio must be.

\item If $n$,  $K$ and $\kappa_d$ satisfy these necessary
    conditions then bistability is only possible if $\kappa_d
    \in [\kappa_{d-}, \kappa_{d+}]$ (c.f. Figure
    \ref{fig:contour-deter}).
\item The locations of the minimal $(x_-)$ and maximal $(x_+)$
    values of $x$ bounding the bistable region  are
    independent of $\kappa_d$.
\item Finally
\begin{enumerate}
\item $(x_+ - x_-)$ is a decreasing function of increasing
    $n$ for constant $\kappa_d, K$
\item $(x_+ - x_-)$ is an increasing function of
    increasing $K$ for constant $n,\kappa_d$.
\end{enumerate}
\end{enumerate}
\end{rem}

\noindent{\bf Local and global stability.} The local stability of
a steady state $x^*$ is determined by the solutions of the
eigenvalue equation \citep{yildirim04}
    \be
    (\lambda + \gamma_1)(\lambda + \gamma_2)(\lambda + \gamma_3) -
    \gamma_1 \gamma_2 \gamma_3 \kappa_d f'_*=0, \quad f'_* = f'(x^*).
    \label{eq:eigenvalue}
    \ee
Set
$$
a_1=\sum_{i=1}^3 \gamma_i, \quad a_2 = \sum_{i\ne j=1}^3 \gamma_i\gamma_j, \quad a_3 = (1-\kappa_d
f'_*)\prod_{i=1}^3 \gamma_i,
$$
so (\ref{eq:eigenvalue}) can be written as
    \be
    \lambda^3 + a_1 \lambda^2 + a_2 \lambda + a_3 = 0.
    \label{eq:eigenvalue1}
    \ee
By Descartes's rule of signs, (\ref{eq:eigenvalue1}) will have
either no positive roots for $f'_* \in [0,\kappa_d^{-1})$ or one
positive root otherwise. With this information  and using the
notation SN to denote a locally stable node, HS a half or
neutrally stable steady state, and US an unstable steady state
(saddle point), then  there will be:
\begin{itemize}
    \item A single steady state $X_1^*$ (SN), for $ \kappa_d
        \in [0, \kappa_{d-})$
    \item Two coexisting steady states $X_1^*$ (SN) and
        $X_2^*=X_3^*$ (HS, born through a saddle node
        bifurcation) for $\kappa_d = \kappa_{d-}$
    \item Three coexisting steady states $X_1^* (SN), X_2^*
        (US), X_3^*$ (SN) for $\kappa_d \in
        (\kappa_{d-},\kappa_{d+})$
    \item Two coexisting steady states $X_1^*=X_2^*$ (HS at a
        saddle node bifurcation), and $X_3^*$ (SN)  for
        $\kappa_d = \kappa_{d+}$
    \item One steady state $X_3^*$ (SN) for $\kappa_{d+} <
        \kappa_d$.
\end{itemize}

For the inducible operon, other work extends these local
stability considerations and we have the following result
characterizing the global behaviour:
\begin{thm}\citep[Proposition 2.1, Chapter 4]{othmer76,smith95}
For an inducible operon with $\varphi$ given by Equation
\ref{ind-phi}, define $I_I = [1/K,1]$.  There is an attracting box
$B_I \subset \mathbb{R}_3^+$ defined by
    \begin{equation*}
    B_I = \{(x_1,x_2,x_3): x_i \in I_I, \,\,i=1,2,3\}
    \end{equation*}
such that the flow $S_t$ is directed inward everywhere
 on the surface of $B_I$. Furthermore, all $X^*
    \in B_I$ and

    \begin{enumerate}
    \item If there is a single steady state, i.e. $X_1^*$ for
        $\kappa_d \in [0,\kappa_{d-})$, or $X_3^*$ for
        $\kappa_{d+} < \kappa_d$, then it is globally stable.
    \item If there are two locally stable nodes, i.e. $X_1^*$
        and  $X_3^*$ for $\kappa_d \in
        (\kappa_{d-},\kappa_{d+})$, then all flows $S(X^0)$
        are attracted to one of them. (See \cite{selgrade79}
        for a delineation of the basin of attraction of
        $X_1^*$ and $X_3^*$.)
    \end{enumerate}
\end{thm}

\subsubsection{Repressible regulation}\label{sss:repress-deter}
\begin{figure}
\centering
\includegraphics[width=\columnwidth]{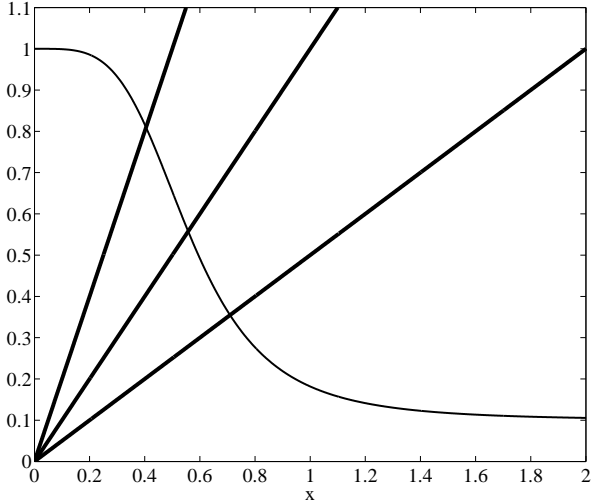}
\caption{Schematic illustration that there is only a single solution of Equation
\ref{eq:ss-dimen} for all values of $\kappa_d$ with repressible
regulation. The monotone decreasing graph is $f$ for a repressible operon,
while the straight lines are $x/\kappa_d$.
This figure was constructed with $n=4$ and $\Delta=10$. See the text for further details.}
\label{fig:1-repressible}
\end{figure}

As illustrated in Figure \ref{fig:1-repressible}, the repressible
operon has a single steady state corresponding to the unique
solution  $x^*$ of Equation \ref{eq:ss-dimen}.  To  determine its
local stability we apply the Routh-Hurwitz criterion to the
eigenvalue equation (\ref{eq:eigenvalue1}). The  steady state
corresponding to $x^*$ will be locally stable (i.e. have
eigenvalues with negative real parts) if and only if $a_1 > 0$
(always the case) and
    \be
    a_1 a_2 - a_3 > 0.
    \label{eq:RH1}
    \ee
The well known relation between the arithmetic and geometric means
$$
\dfrac 1n \sum_{i=1}^n \gamma _i \geq \left( \prod _{i=1}^n \gamma i\right)^{1/n},
$$
when applied to both $a_1$ and $a_2$ gives, in conjunction with
Equation \ref{eq:RH1},
$$
a_1a_2-a_3 \geq  (8+\kappa_d f'_*)  \prod _{i=1}^3 \gamma_i > 0.
$$
Thus as long as $f'_*  > -8/\kappa_d$, the steady state corresponding to $x^*$ will be locally stable. Once
condition \eqref{eq:RH1} is violated, stability of $x^*$ is lost via a supercritical Hopf bifurcation and a
limit cycle is born.  One may even compute the Hopf period of this limit cycle by assuming that $\lambda = j
\omega_H$ ($j=\sqrt{-1}$) in Equation \ref{eq:eigenvalue1} where $\omega_H$ is the Hopf angular frequency.
Equating real and imaginary parts of the resultant yields $\omega_H = \sqrt{a_3/a_1}$ or
    \begin{equation*}
    T_H = \dfrac {2\pi}{\omega_H} =
    {2 \pi}
    \times
    \sqrt {
    \dfrac
    {\sum_{i=1}^3 \gamma_i}
    {(1-\kappa_d f'_*)\prod_{i=1}^3 \gamma_i}
    }.
    \end{equation*}

These local stability results tell us nothing about the global
behaviour when stability is lost, but it is possible to
characterize the global behaviour of a repressible operon with the
following

\begin{thm}\citep[Theorem 4.1 \& Theorem 4.2, Chapter
3]{smith95} For a repressible operon with $\varphi$ given by
Equation \ref{rep-phi}, define $I_R = [K_1/K,1]$. There  is a
globally attracting box $B_R \subset \mathbb{R}_3^+$ defined by
    \begin{equation*}
    B_R = \{(x_1,x_2,x_3): x_i \in I_R, \,\,i=1,2,3\}
    \end{equation*}
such that the flow $S$ is directed inward everywhere on the surface of $B_R$.  Furthermore there is a single
steady state $X^* \in B_R$.  If $X^*$ is locally stable  it is globally stable, but if $X^*$ is unstable then a
generalization  of the Poincare-Bendixson theorem \cite[Chapter 3]{smith95} implies the existence of a globally
stable limit cycle in $B_R$.
\end{thm}

\begin{rem}
There is no necessary connection between the Hopf period computed
from the local stability analysis and the period of the globally
stable limit cycle.
\end{rem}



\section{Fast and slow variables}\label{ssec:fast-slow}

In dynamical systems, considerable simplification and insight into
the behaviour can be obtained by identifying fast and slow
variables. This technique is especially useful when one is
initially interested in the approach to a steady state. In this
context a fast variable is one that relaxes much more rapidly to
an equilibrium than a slow variable \citep{haken83}.  In many
systems, including chemical and biochemical ones, this is often
a consequence of differences in degradation rates, with
the fastest variable the one that has the largest degradation
rate. We employ the same strategy here to obtain approximations to
the population level dynamics that will be used in the next
section.

It is often  the case that the degradation rate of mRNA is much
greater than the corresponding degradation rates for both the
intermediate protein and the effector $(\gamma_1 \gg
\gamma_2,\gamma_3)$ so in this case the mRNA dynamics are fast and
we have the approximate relationship
    \begin{equation*}
    x_1 \simeq \kappa_d f( x_3).
    \end{equation*}
Consequently the three variable system describing the generic
operon reduces to a two variable one involving the slower
intermediate and effector:
    \begin{align}
    \dfrac{dx_2}{dt} &= \gamma_2[\kappa_d f(x_3)-x_2], \label{eq:2intermed}\\
    \dfrac{dx_3}{dt} &= \gamma_3 (x_2 - x_3).\label{eq:2effector}
    \end{align}

In our considerations of specific single operon dynamics below we
will also have occasion to examine two  further subcases, namely

\noindent {\bf Case 1. Intermediate (protein)  dominated
dynamics.}
  If it should happen that
    $\gamma_1 \gg \gamma_3 \gg \gamma_2$ (as for the {\it lac} operon, 
    then
    the effector also qualifies as a fast variable so
    \begin{equation*}
    x_3 \simeq x_2
    \end{equation*}
    and thus from (\ref{eq:2intermed})-(\ref{eq:2effector}) we recover the one dimensional
    equation for the slowest variable, the intermediate:
    \be
    \dfrac{dx_2}{dt} = \gamma_2 [\kappa_d f(x_2) - x_2].
    \label{eq:1D-I}
    \ee

\noindent {\bf Case 2.  Effector (enzyme) dominated dynamics.}
Alternately, if $\gamma_1 \gg \gamma_2 \gg \gamma_3$  then the
intermediate is a fast variable relative to the effector and we
have
    \begin{equation*}
    x_2 \simeq  x_3
    \end{equation*}
so our two variable system (
\ref{eq:2intermed})-(\ref{eq:2effector}) reduces to a one
dimensional system
    \be \dfrac{dx_3}{dt} = \gamma_3[\kappa_d f(x_3) - x_3]
    \label{eq:1D-E}
    \ee
for the relatively slow effector dynamics.

Both Equations \ref{eq:1D-I} and  \ref{eq:1D-E}  are of the form
    \be
   \dfrac{dx}{dt} = \gamma[\kappa_d f(x) -  x]
    \label{eq:1D-general}
    \ee
where $\gamma  $ is either $\gamma_2 $ for protein ($x_2$)
dominated dynamics    or $\gamma_3$ for effector ($x_3$) dominated
dynamics.

\section{Distributions with intrinsic bursting}\label{sec:dynamics-single}

\subsection{Generalities} It is well documented experimentally
\citep{cai,chubb,golding,raj,sigal,yu} that in some organisms the
{\it amplitude} of protein production through bursting translation
of mRNA is exponentially distributed at the single cell level with
density
    \be
    h(y) = \dfrac 1 {\bar b} e^{-y/{\bar b}},
    \label{eq:bursting-den}
    \ee
where $\bar b$ is the average burst size, and that the frequency
of bursting $\varphi$ is dependent on the level of the effector.
Writing Equation \ref{eq:bursting-den} in terms of our
dimensionless variables we have
    \be
    h(x) = \dfrac 1 {b} e^{-x/{b}}.
    \label{eq:bursting-den-dimen}
    \ee

\begin{rem}

The technique of eliminating fast variables described in Section
\ref{ssec:dynamics-ensemble} above (also known as the adiabatic
elimination technique \citep{haken83}) has been extended to
stochastically perturbed systems when the perturbation is a
Gaussian distributed white noise, c.f. \citet[Chapter 4, Section
11.1]{stratonovich-vol1}, \citet{wilemski76}, \citet{titulaer78},
and \citet[Section 6.4]{gardinerhandbook}.  However, to the best
of our knowledge, this type of approximation has never been
extended to the situation dealt with here in which the
perturbation is a jump Markov process.
\end{rem}

The  single cell analog of the population level intermediate
protein dominated Case 1 above (when $\gamma_1 \gg \gamma_3 \gg
\gamma_2$) is
    \be
    \dfrac{dx_2}{dt} = -\gamma_2 x_2 + \Xi (h,\varphi(x_2) ),
    \quad \mbox{with} \quad \varphi(x_2) = \gamma_2 \varphi_m
    f(x_2),
    \label{eq:single-case1}
    \ee
where $\Xi (h,\varphi)$ denotes a jump  Markov process, occurring
at a rate $\varphi$, whose amplitude is distributed with density
$h$ as given in (\ref{eq:bursting-den-dimen}). Analogously, in the
Case 2 effector dominated situation the single cell equation
becomes
    \be
    \dfrac{dx_3}{dt} = -\gamma_3 x_3 + \Xi (h, \varphi(x_3) ),
    \quad \mbox{with} \quad \varphi(x_3) = \gamma_3 \varphi_m
    f(x_3).
    \label{eq:single-case2}
    \ee
Equations \ref{eq:single-case1} and \ref{eq:single-case2} can both
be written as
 \begin{equation*}
    \dfrac {dx}{dt} = -\gamma x + \Xi (h,\varphi(x)),
    \quad \mbox{with} \quad \varphi(x) =
    \gamma \kappa_b f(x), \,\, \kappa_b \equiv \varphi_m.
    \label{eq:single}
    \end{equation*}
\begin{rem} In the case of bursting we will always take $\kappa_b
\equiv \varphi_m$ in contrast to the deterministic case where
$\kappa_d = b_d \varphi_m$.
\end{rem}

From \citet{mackeytyran08}  the corresponding operator
equation for the evolution of the density $u(t,x)$  when there is
a single dominant slow variable is given by
    \begin{equation}
    \begin{split}
    \dfrac{\partial u(t,x)}{\partial t} -\gamma \dfrac{\partial (xu(t,x))}{\partial
    x}& = -\gamma \kappa_b f(x)  u(t,x)\\
    & \quad + \gamma \kappa_b \int_0^x  f(  y) u(t,y) h(x-y)dy.
    \label{eq:operator-eqn}
    \end{split}
    \end{equation}
\begin{rem}
This is a straightforward generalization of what \citet[Section
3.4]{gardinerhandbook} refers to as the differential
Chapman-Kolmogorov equation.
\end{rem}

Stationary solutions $u_*(x)$ of (\ref{eq:operator-eqn}) are
solutions of \be
    - \dfrac{d (xu_*(x))}{d
    x} = -\kappa_b f(  x)u_*(x) + \kappa_b \int_0^x  f(  y) u_*(y) h(x-y)dy.
    \label{eq:ss-operator-eqn}
    \ee
If there is a unique stationary density, then the solution $u(t,x)$ of Equation \ref{eq:operator-eqn} is said to
be asymptotically stable \citep{almcmbk94} in the sense that
 $$
    \lim_{t\to \infty} \int_0^\infty |u(t,x) -u_*(x)|dx = 0
$$
for all initial densities $u(0,x)$.

\begin{thm}\cite[Theorem
7]{mackeytyran08}. The unique stationary density of Equation
\ref{eq:ss-operator-eqn}, with $f$ given by Equation
\ref{eq:gen-response-fun-dimen} and $h$ given by
(\ref{eq:bursting-den}), is
    \be
    u_*(x) = \dfrac{\mathcal{C}}{x} e^{-x/b} \exp \left [ \kappa_b
    \int^x \frac{f(  y)}{y}dy \right
    ],
    \label{eq:ss-soln}
    \ee
where $\mathcal{C}$ is a normalizing constant such that
$\int_0^\infty u_*(x)dx = 1$.  Further, $u(t,x)$ is asymptotically
stable.
\end{thm}

\begin{rem}\label{r:invd}
The stationary density (\ref{eq:ss-soln}) is found by rewriting
Equation \ref{eq:ss-operator-eqn} in the form
    $$
    \dfrac{d y(x)}{dx} + \dfrac{ y(x)}{b} -\kappa_b \dfrac{f(x)}{x}y(x)=0, \quad y(x) \equiv xu_*(x)
    $$
using Laplace transforms and solving by quadratures. Note also that we can represent $u_*$ as
\[
u_*(x)=\mathcal{C}\exp\int^x\left(\frac{\kappa_b f(y)}{y}-\frac{1}{b}-\frac{1}{y}\right)dy,
\]
where $\mathcal{C}$ is a normalizing constant.
\end{rem}

\subsection{Distributions in the presence of bursting}\label{sssec:induc-repress}

\subsubsection{Protein distribution in the absence of
control}\label{sssec:noctl}

If the burst frequency $\varphi = \gamma \kappa_b f $ is
independent of the level of all of the participating molecular
species, then  the solution given in Equation \ref{eq:ss-soln} is
the density of the gamma distribution:
    \begin{equation*}
    u_*(x) = \dfrac{1}{b^{\kappa_b } \Gamma(\kappa_b)} x^{\kappa_b  -1 }
    e^{-x/b},
    \label{eq:gamma}
    \end{equation*}
where $\Gamma(\cdot)$ denotes the gamma function. For $\kappa_b \in (0,1)$, $u_*(0)=\infty$ and $u_*$ is
decreasing while for $\kappa_b
>1$, $u_*(0)=0$ and  there is a maximum at $x=b(\kappa_b -1)$.

\subsubsection{Controlled bursting}

We next consider the situation in which the burst frequency
$\varphi$ is dependent on the level of $x$, c.f. Equation
\ref{eq:gen-response-fun}.  This requires that we evaluate
\begin{equation*}
\begin{split}
    \kappa_b
    \int^x \dfrac{f(y)}{y}dy
    =  \int^x \dfrac {\kappa_b}{y} \left [ \dfrac{1+  y^n}{\Lambda + \Delta y^n}  \right
    ]dy
    = \ln \left \{ x^{\kappa_b \Lambda^{-1} } (\Lambda+ \Delta x^n)^\theta \right \},
    \end{split}
    \end{equation*}
where $\Lambda, \Delta$ are enumerated in Table \ref{tab:ABC1} for both the inducible and repressible operons
treated in Section \ref{ssec:response} and
\[
\theta=\frac{\kappa_b}{n\Delta}\left(1-\frac{\Delta}{\Lambda}\right).
\]
Consequently, the steady state density (\ref{eq:ss-soln}) explicitly becomes
    \be
    u_*(x) = {\mathcal{C}}e^{-x/b} x^{\kappa_b \Lambda^{-1}  -1 } (\Lambda + \Delta x^n)^\theta.
    \label{eq:ss-soln-integ}
    \ee

The first two terms of Equation \ref{eq:ss-soln-integ} are simply proportional to the density of the gamma
distribution. For $0 < \kappa_b \Lambda^{-1} < 1 $ we have $u_*(0)=\infty$ while for $\kappa_b \Lambda^{-1}
> 1 $, $u_*(0)=0$ and  there is at least one maximum at a value of $x
> 0$. We have $u_*(x)>0$ for all $x>0$ and from Remark~\ref{r:invd} it follows that
 \begin{equation}\label{eq:ss-soln-integ-deriv1}
u_*'(x)=u_*(x)\left(\frac{\kappa_b f(x)}{x}-\frac{1}{b}-\frac{1}{x}\right), \quad x>0.
 \end{equation}
Observe that if $\kappa_b\le 1$ then  $u_*$ is a monotone decreasing function of $x$, since $\kappa_b f(x)\le 1
$ for all $x>0$. Thus we assume in what follows that $\kappa_b>1$.

Since the analysis of the qualitative nature of the stationary
density leads to different conclusions for the inducible and
repressible operon cases, we consider each in turn.

\subsubsection{Bursting in the inducible
operon}\label{sss:inducible-dist}

\begin{figure}
\centering
\includegraphics[width=\columnwidth]{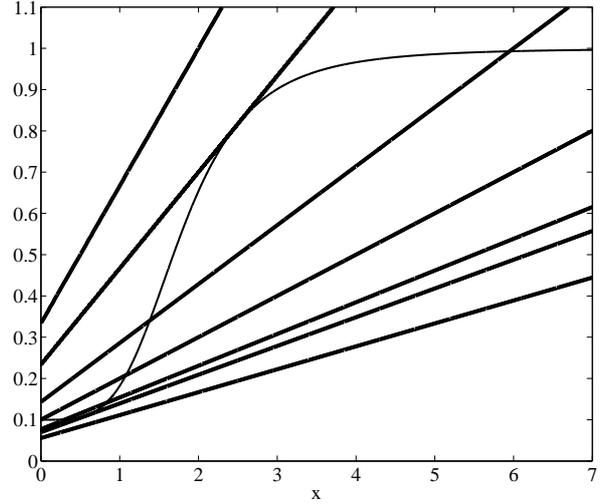}
\caption{ Schematic illustration of the possibility of one, two or three solutions of Equation \ref{eq:zero} for
varying values of $\kappa_b$ with bursting inducible regulation. The straight lines correspond (in a clockwise
direction) to $\kappa_b \in (0,\kappa_{b-})$, $\kappa_b = \kappa_{b-}$, $\kappa_b \in (\kappa_{b-},\kappa_{b+})$
(and respectively $\kappa_b<K$, $\kappa_b=K$, $K<\kappa_b$), $\kappa_b = \kappa_{b+}$, and $\kappa_{b+} <
\kappa_b$. This figure was constructed with $n=4$, $K=10$ and $b=1$ for which $\kappa_{b-}=4.29$ and
$\kappa_{b+} = 14.35$ as computed from (\ref{eq:kappapm-burst}).  See the text for further details.
}
\label{fig:123extremum}
\end{figure}

\begin{figure}
\centering
\includegraphics[width=\columnwidth]{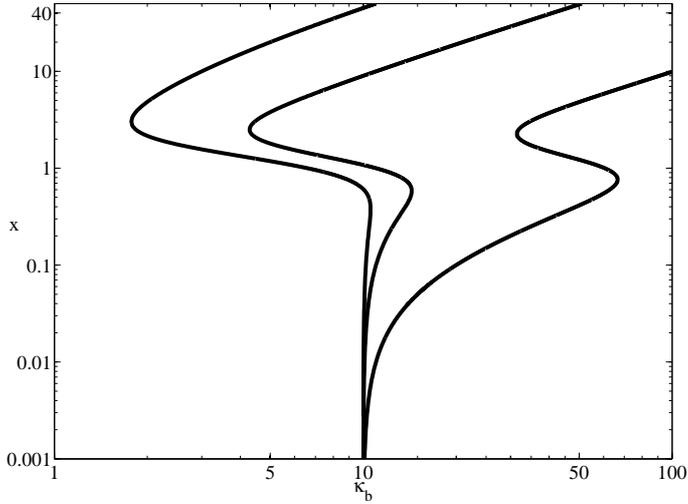}
\caption{Full logarithmic plot of the values of $x$ at which $u_*'(x) = 0$ versus the parameter
$\kappa_b$, obtained from Equation \ref{eq:zero},
for $n=4$, $K = 10$, and (left to right) $b=5, 1$ and $b= \frac 1{10}$.   Though somewhat obscured by
the logarithmic scale for $x$, the graphs always intersect the $\kappa_b$ axis at $\kappa_b = K$. Additionally, it
is important to note that $u'_*(0)=0$ for $K<\kappa_b$,
and that there is always a maximum at $0$ for $0<\kappa_b<K$. See the text for further details.  }
\label{fig:burst-inducible-1}
\end{figure}

For $\theta> 0$, as in the case of an inducible operon, the third term of Equation \ref{eq:ss-soln-integ} is a
monotone increasing function of $x$ and, consequently, there is the possibility that $u_*$ may have more than
one maximum,  indicative of the existence of bistable behaviour. In this case, the stationary density becomes
    \begin{equation*}
    u_*(x) = {\mathcal{C}}e^{-x/b} x^{ \kappa_b K^{-1}  -1 } (K+x^n)^\theta, \quad \theta = \dfrac {\kappa_b}{
    n}
    (1-K^{-1}).
    \label{eq:ss-soln-induc}
    \end{equation*}
From \eqref{eq:ss-soln-integ-deriv1} it follows that we have  $u_*'(x)=0$ for $x>0$ if and only if
    \be
    \dfrac 1{\kappa_b } \left ( \dfrac x b + 1\right ) = \dfrac {1 + x^n}{K +
    x^n}.
    \label{eq:zero}
    \ee
Again, graphical arguments (see Figure \ref{fig:123extremum}) show that there may be up to three roots of
(\ref{eq:zero}).
For illustrative values of $n$, $K$, and $b$, Figure
\ref{fig:burst-inducible-1} shows the graph of the values of $x$
at which $u_*'(x)=0$ as a function of $\kappa_b$. 
When there are three roots of
(\ref{eq:zero}), we label them as
$\tilde x_1 < \tilde x_2 < \tilde x_3$.

Generally  we cannot determine when there are three roots.
However, we can determine when there are only two roots $\tilde
x_1 < \tilde x_3$ from the argument of Section
\ref{sss:induc-deter}. At $\tilde x_1$ and $ \tilde x_3$ we will
not only have Equation \ref{eq:zero} satisfied but the graph of
the right hand side of (\ref{eq:zero}) will be tangent to the
graph of the left hand side at one of them so the slopes will be
equal. Differentiation of (\ref{eq:zero}) yields the second
condition
    \be
        n \dfrac {x^{n-1}}{(K+x^n)^2} = 
        \dfrac {1}{\kappa_b b (K-1 )}
    \label{eq:burst-tangency1}
    \ee

%
%

We first show that there is an open set of parameters $(b,K,\kappa_b)$ for which the stationary density $u_*$ is
bimodal. From Equations \ref{eq:zero} and \ref{eq:burst-tangency1} it follows that the value of $x_{\pm}$ at
which tangency will occur is given by
\[
x_\pm=b(\kappa_b-1)z_\pm
\]
and $z_\pm$ are positive solutions of equation
\begin{equation*}
\frac{z}{n}=1-z-\beta(1-z)^2, \quad \text{where}\quad \beta=\frac{K(\kappa_b-1)}{(K-1)\kappa_b}.
\end{equation*}
We explicitly have
\begin{equation*}
z_\pm=\frac{1}{2\beta n}\left(2\beta n-\left(n+1\right)\pm\sqrt{\left(n+1\right)^2-4\beta n}\right)
\end{equation*}
provided that
\begin{equation}\label{e:delta}
\frac{(n+1)^2}{4n}\ge \beta=\frac{K(\kappa_b-1)}{(K-1)\kappa_b}.
\end{equation}
Equation \ref{e:delta} is always satisfied when $\kappa_b<K$ or
when $\kappa_b>K$ and $K$ is as in the deterministic case
\eqref{eq:K-necessary}. Observe also that we have $z_+>0>z_{-}$
for $\kappa_b<K$ and $z_+>z_{-}>0$ for $\kappa_b>K$. The two
corresponding values of $b$ at which a tangency occurs are given
by
\[
b_\pm=\frac{1}{(\kappa_b-1)z_\pm}\sqrt[n]{\frac{K}{\beta(1-z_\pm)}-K}\quad\text{and}\quad z_{\pm}>0.
\]
If $\kappa_b<K$ then $u_*(0)=\infty$ and $u_*$ is decreasing for $b\le b_+$, while for $b>b_+$ there is a local
maximum at $x>0$. If $\kappa_b>K$ then $u_*(0)=0$ and $u_*$ has one or two local maximum. As a consequence, for
$n>1$ we have a bimodal steady state density $u_*$ if and only if the parameters $\kappa_b$ and $K$
 satisfy \eqref{e:delta}, $\kappa_b>K$, and $b\in (b_+,b_{-})$.

We now want to find the analogy between the bistable behavior in the deterministic system and the existence of
bimodal stationary density $u_*$. To this end we fix the parameters $b>0$ and $K>1$ and vary $\kappa_b$ as in
Figure~\ref{fig:123extremum}. Equations \ref{eq:zero} and \ref{eq:burst-tangency1} can also be combined to give
an implicit equation for the value of $x_{\pm}$ at which tangency will occur
    \be
    x^{2n} - (K-1)\left [ n - \dfrac{K+1}{K-1}\right ]x^n -
nb(K-1)x^{n-1} +K=0
    \label{eq:xpm-implicit}
    \ee
and the corresponding values of $\kappa_{b \pm}$ are given by
    \be
    \kappa_{b \pm} = \left (\dfrac{x_{\mp} +b}{b} \right )
    \left(
    \dfrac{K+x_{\mp}^n}{1+x_{\mp}^n}
    \right).
    \label{eq:kappapm-burst}
    \ee
There are two cases to distinguish.

\noindent {\bf Case 1. $0<  \kappa_b < K$.}  In this case, $u_*(0)=\infty$. Further, the same graphical
considerations as in the deterministic case show that there can be none, one, or two positive solutions to
Equation \ref{eq:zero}. If $\kappa_b < \kappa_{b-}$, there are no positive solutions, $u_* $ is a monotone
decreasing function of $x$. If $\kappa_b > \kappa_{b-}$, there are two positive solutions ($\tilde x_2$ and
$\tilde x_3$ in our previous notation, $\tilde x_1$ has become negative and not of importance) and there will be
a maximum in $u_* $ at $\tilde x_3$ with a minimum in $u_* $ at $\tilde x_2$. 

\noindent {\bf Case 2. $0 < K < \kappa_b $.}  Now, $u_*(0)=0$ and   there may be one, two, or three positive
roots of Equation \ref{eq:zero}. We are interested in knowing when there are three which we label as $\tilde x_1
< \tilde x_2 < \tilde x_3$ as $\tilde x_1, \tilde x_3$ will correspond to the location of maxima in $u_* $ while
$\tilde x_2$ will be the location of the minimum between them and the condition for the existence of three roots
is $\kappa_{b-} < \kappa_{b} < \kappa_{b+}$.

We see then that the different possibilities depend on the respective values of $K$, $\kappa_{b-}$,
$\kappa_{b+}$, and $\kappa_{b}$. To summarize, we may characterize the stationary density $u_*$ for an inducible
operon in the following way:
\begin{enumerate}
\item {\bf Unimodal type 1}: $u_*(0)=\infty$ and $u_*$ is decreasing for
    $0 < \kappa_{b} < \kappa_{b-}$ and $0 < \kappa_{b} < K$
\item {\bf Unimodal type 2}: $u_*(0)=0$ and $u_*$ has a single
    maximum~at
\begin{enumerate}
    \item $\tilde x_1> 0$ for $ K < \kappa_b <
        \kappa_{b-} $ or
    \item at $\tilde x_3> 0$ for $\kappa_{b+} <
        \kappa_b $ and $ K <  \kappa_b$
    \end{enumerate}
\item {\bf Bimodal type 1}:  $u_*(0)=\infty$ and $u_*$ has
    a single maximum at $\tilde x_3 > 0$ for $\kappa_{b-} <
    \kappa_b < K$
\item {\bf Bimodal type 2}:  $u_*(0)=0$ and $u_*$ has two maxima at $\tilde
    x_1,\tilde x_3$,  $0 < \tilde x_1 < \tilde x_3$ for
    $\kappa_{b-} < \kappa_b < \kappa_{b+}$ and $K<\kappa_b$
\end{enumerate}

\begin{rem}
Remember that the case $n=1$ cannot display   bistability in the deterministic case. However, in the case of
bursting in the inducible system  when $n=1$, if $\frac{K}{b}+1<\kappa_b<K$ and $b>\frac{K}{K-1}$, then
$u_*(0)=\infty$ and $u_*$ also has a maximum at $\tilde x_3 > 0$.  Thus in this case one can have a Bimodal type 1
stationary density.
\end{rem}


\begin{figure}
\centering
\includegraphics[width=\columnwidth]{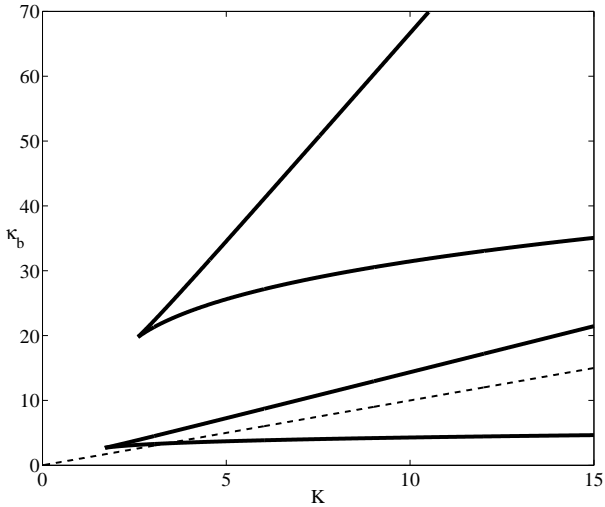}
\caption{In this figure we present two  bifurcation diagrams (for $n=4$) in $(K,\kappa_b)$ parameter
space delineating unimodal from bimodal stationary densities $u_*$ in an inducible operon with bursting
as obtained from Equations \ref{eq:burst-induc-para} with \ref{eq:F}.  The upper cone-shaped
plot is for $b=\frac 1{10}$ while the bottom one is for $b=1$.  In both cone shaped regions,
for any situation in which the lower branch is above the line $\kappa_b = K$ (lower straight line) then bimodal behaviour
in the stationary solution $u_*(x)$ will be observed with maxima in $u_*$ at positive values of $x$, $\tilde x_1$ and $\tilde x_3$.
}
\label{fig:contour-inducible-burst}
\end{figure}

We now choose to see how the average burst size $b$ affects bistability in the density $u_*$ by looking at the
parametric plot of $\kappa_b (x)$ versus $K(x)$. Define
    \be
    F(x,b) = \dfrac {x^n+1}{nx^{n-1}(x+b)}.
    \label{eq:F}
    \ee
Then
    \be K(x,b)=\dfrac {1+x^nF(x,b)}{1-F(x,b)} \quad
    \mbox{and}\quad \kappa_b (x,b) = [K(x,b)+x^n] \dfrac
    {x+b}{b(x^n+1)}.
    \label{eq:burst-induc-para}
    \ee
The bifurcation diagram obtained from a parametric plot of $K$
versus $\kappa_b$ (with $x$ as the parameter) is illustrated in
Figure \ref{fig:contour-inducible-burst} for $n=4$ and two values
of $b$.  Note that it is necessary for $0 < K <\kappa_b$ in order
to obtain Bimodal type 2 behaviour.

For bursting behaviour in an inducible situation, there are two different bifurcation patterns that are
possible.  The two different cases are delineated by the respective values of $K$ and $\kappa_{b}$, as  shown in
Figure \ref{fig:burst-inducible-1} and Figure \ref{fig:contour-inducible-burst}. Both bifurcation scenarios
share the property that while increasing the bifurcation parameter $\kappa_b$ from $0$ to $\infty$, the
stationary density $u_*$ passes from a unimodal density with a peak at a low value (either $0$ or $\tilde x_1$)
to a bimodal density and then back to a unimodal density with a peak at a high value ($\tilde x_3$).

In what will be referred as \textbf{Bifurcation type 1}, the
maximum at $x=0$ disappears when there is a second peak at
$x=\tilde x_3$. The sequence of densities encountered for
increasing values of $\kappa_b$ is then:   Unimodal type 1 to a
Bimodal type 1 to a Bimodal type 2 and finally to a Unimodal type
2 density.
\begin{figure}
\centering
\includegraphics[width=\columnwidth]{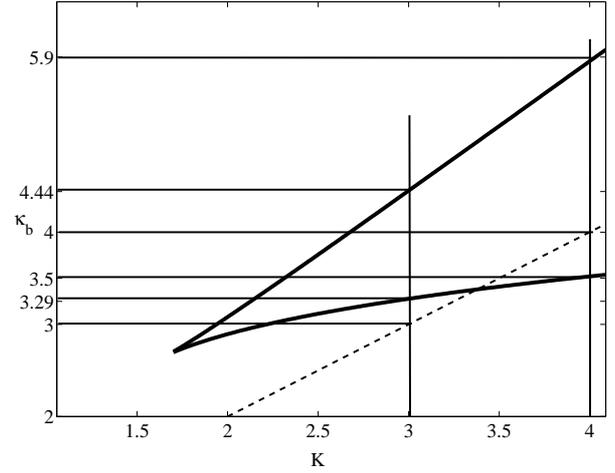}
\caption{This figure presents an enlarged portion of Figure \ref{fig:contour-inducible-burst} for $b=1$.  The
various horizontal lines mark specific values of $\kappa_b$ referred to in Figures \ref{fig:intrinsic_bif1} and
\ref{fig:intrinsic_bif2}.} \label{fig:contour-inducible-burst-zoom}
\end{figure}

In the \textbf{Bifurcation type 2} situation, the sequence of
density types for increasing values of $\kappa_b$ is:  Unimodal
type 1  to a Unimodal type 2 and then a Bimodal type 2 ending in a
Unimodal type 2 density.

\begin{figure}[tb]
\centering
\includegraphics[width=\columnwidth]{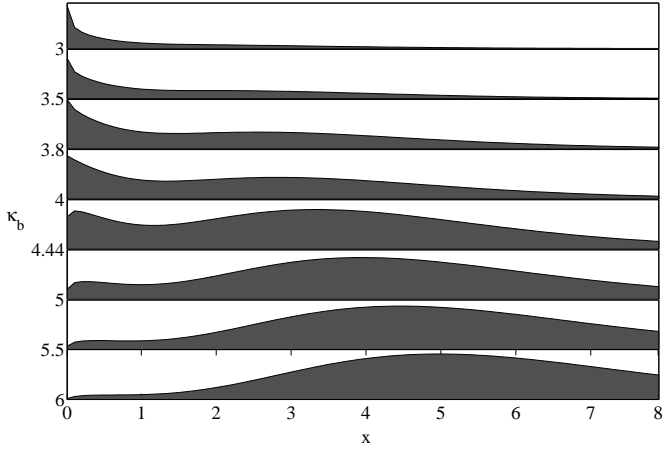}
\caption{In this figure we illustrate {\bf Bifurcation type 1} when intrinsic bursting is present. For a variety
of values of the bifurcation parameter $\kappa_b$ (between $3$ and $6$ from top to down), the stationary density $u_*$ is
plotted versus $x$ between $0$ and $8$. The values of the parameters used in this figure are $b=1$, $K=4$, and
$n=4$. For $\kappa_b \lesssim 3.5 $, $u_*$ has a single maximum at $x=0$. For $3.5 \lesssim \kappa_b < 4 $,
$u_*$ has two local maxima at $x=0$ and $\tilde x_3 >1$. For $4 < \kappa_b \lesssim 5.9 $, $u_*$ has two local
maxima at $0<\tilde x_1<\tilde x_3$. Finally, for $\kappa_b \gtrsim 5.9 $, $u_*$ has a single maximum at $\tilde
x_3 >1$. Note that for each plot of the density, the scale of the ordinate is arbitrary to improve the visualization.} \label{fig:intrinsic_bif1}
\end{figure}

The  two different kinds of bifurcation that can occur are easily illustrated for $b=1$ as the parameter
$\kappa_b$ is increased.  (An enlarged diagram in the region of interest is shown in Figure
\ref{fig:contour-inducible-burst-zoom}.) Figure \ref{fig:intrinsic_bif1} illustrates  {\bf Bifurcation type 1},
when $K=4$, and $\kappa_b$ increases from low to high values. As $\kappa_b$ increases, we pass from a Unimodal
type 1 density, to a Bimodal type 1 density. Further increases in  $\kappa_b$ lead to a Bimodal type 2 density
and finally to a Unimodal type 2 density. This bifurcation cannot occur, for example, when $b=\frac {1}{10}$ and
$K \leq 15$ (see Figure \ref{fig:contour-inducible-burst}).

\begin{figure}[tb]
\centering
\includegraphics[width=\columnwidth,height=5.96cm]{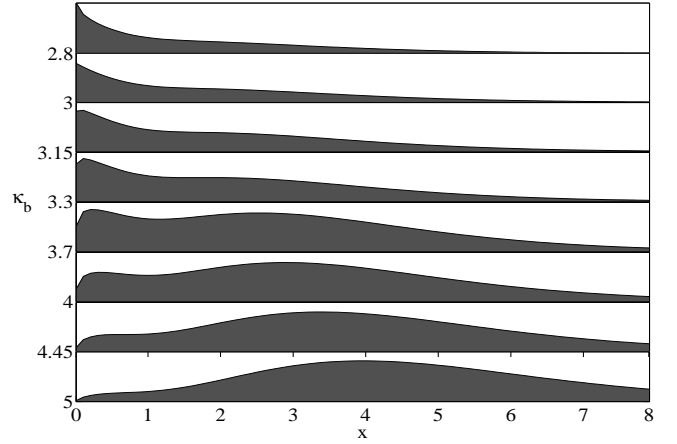}
\caption{An illustration of {\bf Bifurcation type 2} for intrinsic bursting. For several values of the
bifurcation parameter $\kappa_b$ (between $2.8$ and $5$ from top to down), the stationary density $u_*$ is plotted versus $x$
between $0$ and $8$. The parameters used are $b=1$, $K=3$, and $n=4$. For $\kappa_b <  3 $, $u_*$ has a single
maximum at $x=0$, and for $3 < \kappa_b \lesssim 3.3 $, $u_*$ has a single maximum at $\tilde x_1 >0$. For $3.3
\lesssim \kappa_b \lesssim 4.45 $, $u_*$ has two local maxima at $0 < \tilde x_1 < \tilde x_3$, and finally for
$\kappa_b \gtrsim 4.45 $ $u_*$ has a single maximum at $\tilde x_3 >0$. Note that for each plot of the density, the scale of the ordinate is abritrary to improve the visualization.    } \label{fig:intrinsic_bif2}
\end{figure}

Figure \ref{fig:intrinsic_bif2} shows  {\bf Bifurcation type 2},
when $K=3$.  As $\kappa_b$ increases, we pass from a Unimodal type
1 density, to a Unimodal type 2 density. Then with further
increases in $\kappa_b$,  we pass to a Bimodal type 2 density and
finally back to a Unimodal type 2 density.

\begin{figure}[htb]
\centering
\includegraphics[width=\columnwidth]{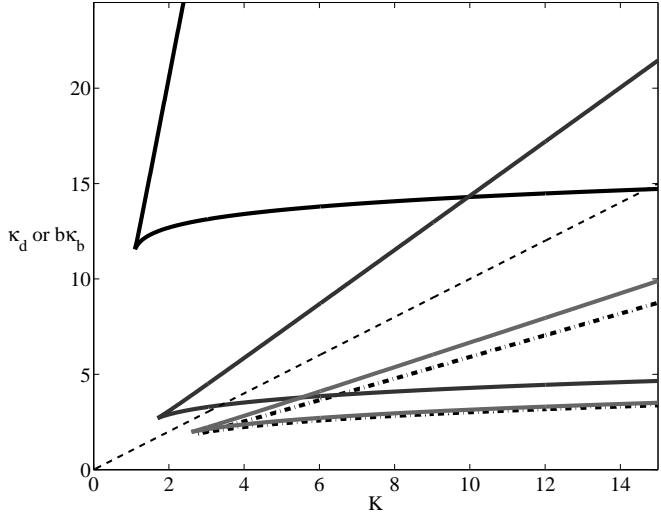}
\caption{The presence of bursting can drastically alter regions of bimodal behaviour as shown in this parametric
plot (for $n=4$) of the boundary in $(K,b \cdot \kappa_b)$ parameter space delineating unimodal from bimodal
stationary densities $u_*$ in an inducible operon with bursting and in $(K,\kappa_d)$ parameter space
delineating one from three steady states in the deterministic inducible operon. From top  to bottom, the regions
are for $b=10$, $b=1$, $b=0.1$ and $b=0.01$. The lowest (heavy dashed line) is for the deterministic case. Note
that for   $b = 0.1$, the two regions of bistability and bimodality coincide and are indistinguishable from one
another.  } \label{fig:alter_bistability}
\end{figure}

\begin{rem} There are several qualitative conclusions to be
drawn from the analysis of this section.
\begin{enumerate}
    \item The presence of bursting can drastically alter the
        regions of parameter space in which bistability can
        occur relative to the deterministic case. Figure
        \ref{fig:alter_bistability} presents  the regions of
        bistability in the presence of bursting in the $(K,b
        \cdot \kappa_b)$ parameter space, which should be
        compared to the region of bistability in the
        deterministic case in the $(K,\kappa_d)$ parameter
        space ($b\kappa_b$ is the mean number of proteins
        produced per unit of time, as is $\kappa_d$)
    \item When $0<\kappa_b <K$, at a fixed value of
        $\kappa_b$, increasing the average burst size $b$ can
        lead to a bifurcation from Unimodal type 1 to Bimodal
        type 1.
    \item When $0<K<\kappa_b$, at a fixed value of $\kappa_b$,
        increasing $b$ can lead to a bifurcation from Unimodal
        type 2 to Bimodal type 2 and then back to Unimodal
        type 2.

\end{enumerate}

\end{rem}

\subsubsection{Bursting in the repressible
operon}\label{sss:repress-dist}

\begin{figure}
\centering
\includegraphics[width=\columnwidth]{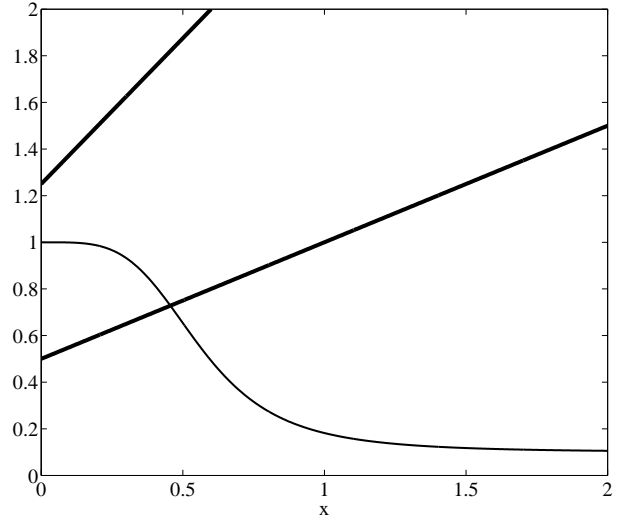}
\caption{Schematic illustration that there can be one or no solution of Equation \ref{eq:zero_repressible},
depending on the value of $\kappa_b$, with repressible regulation. The straight lines correspond (in a clockwise
direction) to $\kappa_b=2$ and $\kappa_b=0.8$. This figure was constructed with $n=4$, $\Delta=10$ and $b=1$.
See the text for further details.} \label{fig:1_bursting_repressible}
\end{figure}

The possible behaviours in the stationary density $u_*$ for the
repressible operon are easy to delineate based on the analysis of
the previous section, with Equation \ref{eq:zero} replaced by
    \be
   \dfrac {1}{\kappa_b} \left ( \dfrac x b + 1\right ) = \dfrac {1 + x^n}{1 +
   \Delta x^n}.
   \label{eq:zero_repressible}
   \ee
Again graphical arguments (see Figure
\ref{fig:1_bursting_repressible})  show that Equation
\ref{eq:zero_repressible} may have either none or one solution.
Namely,

\begin{enumerate}

\item For $0 < \kappa_b < 1$, $u_*(0)=\infty$ and $u_*$ is decreasing. Equation \ref{eq:zero_repressible} does not have
    any solution (Unimodal type 1).
    \item For $1<\kappa_b$, $u_*(0)=0$ and $u_*$ has a single maximum at a
        value of $x >0$ determined by the single positive
        solution of Equation \ref{eq:zero_repressible}
        (Unimodal type 2).

\end{enumerate}

\subsection{Recovering the deterministic case}

We can recover the deterministic behaviour from the bursting dynamics with a suitable scaling of the parameters
and limiting procedure. With  bursting production there are two important parameters (the frequency $\kappa_b$
and the amplitude $b$), while with deterministic production there is only $\kappa_d$. The natural limit to
consider is when
    \begin{equation*}
    b \to 0, \quad \kappa_b \to \infty
    \quad\mbox{with} \quad b \kappa_b \equiv  \kappa_d.
    \label{limitdeterminist_intrinsic}
    \end{equation*}
In this limit,  the implicit  equations
which define
the maximum points of the steady state density, become  the
implicit equations (\ref{eq:ss-dimen}) and (\ref{eq:slope}) which
define the stable steady states in the deterministic case.

The bifurcations will also take place at the same points, because
we recover 
Equation \ref{eq:kappa-pm} in the limit. However, Bimodality type 1 as well as the Unimodal type 1 behaviours
will no longer be present, as in the deterministic case, because for $\kappa_b \to \infty$ we have $\kappa_b >
K$. Finally, from the analytical expression for the steady-state density (\ref{eq:ss-soln-integ}) $u_*$  will
became more sharply peaked as $b \to 0$. Due to the normalization constant (which depends on $b$ and
$\kappa_b$), the mass will be more concentrated around the larger maximum of~$u_*$.

\section{Distributions with fluctuations in the degradation rate}\label{sec:dynamics-degrad}

\subsection{Generalities}

For a generic one dimensional stochastic differential equation of
the form
    \begin{equation*}
    dx(t) = \alpha(x) dt + \sigma(x)dw(t)
    \label{eq:generic-sde}
    \end{equation*}
the corresponding Fokker Planck equation
    \be
    \dfrac
    {\partial u}{\partial t} = - \dfrac {\partial (\alpha u)}{\partial x}
    + \dfrac 12 \dfrac {\partial^2 (\sigma^2u)}{\partial x^2}
    \label{eq:generic-fp}
    \ee
can be written in the form of a conservation equation
    \begin{equation*}
    \dfrac
    {\partial  u}{\partial t} + \dfrac {\partial J}{\partial x} = 0,
    \label{eq:generic-conservation}
    \end{equation*}
where
    \begin{equation*}
    J = \alpha u - \frac 12 \dfrac {\partial (\sigma^2u)}{\partial x}
    \label{eq:current}
    \end{equation*}
is  the probability current.  In a steady state when $\partial_t u
\equiv 0$, the current must satisfy $J = \mbox{constant}$
throughout the domain of the problem.  In the particular case when
$J = 0$ at one of the boundaries (a reflecting boundary) then
$J=0$ for all $x$ in the domain and the steady state solution
$u_*$ of Equation \ref{eq:generic-fp} is easily obtained with a
single quadrature as
    \begin{equation*}
    u_*(x) = \dfrac {\mathcal C}{\sigma^2(x)} \exp
    \left \{
    2 \int^x \dfrac {\alpha (y)}{\sigma^2(y)}dy
    \right \},
    \label{eq:generic-ss}
    \end{equation*}
where $\mathcal C$ is a normalizing constant as before.

\subsection{Fluctuations in degradation rate}

In our considerations of the  effects of continuous fluctuations, we examine the  situation in which
fluctuations appear in the degradation rate $\gamma$ of the generic equation (\ref{eq:1D-general}).  From
standard chemical kinetic arguments \citep{oppenheim69}, if the fluctuations are Gaussian distributed the mean
numbers of molecules decaying in a time $dt$ is simply $\gamma xdt$ and the standard deviation of these numbers
is proportional to $\sqrt{x}$.  Thus we take the decay to be given by the sum of a deterministic component
$\gamma x dt$ and a stochastic component $ \sigma \sqrt{x} dw(t)$, where $w$ is a standard Brownian motion, and
write Equation \ref{eq:1D-general} as a stochastic differential equation in the form
    \begin{equation*}
    dx = \gamma [\kappa_d f (  x)  - x] dt + \sigma \sqrt{x}
    dw.
    \label{eq:1d-stoch-decay}
    \end{equation*}
Within the Ito interpretation of stochastic integration, this equation has a corresponding Fokker Planck
equation for the evolution of the ensemble density $u(t,x)$ given by \citep{almcmbk94}
    \be
     \dfrac
     {\partial u}{\partial t} = - \dfrac {\partial
    \left [
    (\gamma \kappa_d f (  x) - \gamma x) u
    \right ]}{\partial x}
    + \dfrac {\sigma^2}{2} \dfrac {\partial ^2 ( x u )}{\partial x^2}.
    \label{eq:1d-stoch-decay-fp}
    \ee

In the situation we consider here, $\sigma(x) = \sigma \sqrt{x}$
and $\alpha (x) =  \gamma \kappa_d f (  x) - \gamma x$.  Further,
since concentrations of molecules cannot become negative the
boundary at $x=0$ is reflecting and the stationary solution of
Equation \ref{eq:1d-stoch-decay-fp} is given by
    \begin{equation*}
    u_*(x) = \dfrac{\mathcal{C}}{x} e^{-2\gamma x/\sigma^2}
    \exp \left [ \frac {2\gamma \kappa_d }{ \sigma ^2}\int^x \frac{f (y)}{y}dy \right
    ].
    \label{eq:ss-soln-decay}
    \end{equation*}
Set $\kappa_e = {2\gamma \kappa_d  / \sigma ^2}$. Then the steady state solution is given explicitly by
    \be
    u_*(x) = {\mathcal{C}}e^{-2\gamma x/\sigma^2}  x^{ \kappa_e \Lambda^{-1} -1 } [\Lambda +
    \Delta x^n]^\theta,
    \label{eq:ss-soln-integ-decay}
    \ee
where $\Lambda,\Delta \geq 0$ and $\theta$ are given in Table
\ref{tab:ABC1}.
\begin{rem}  Two comments are in order.

\ben \item Because the form of the solutions for the situation
with bursting (intrinsic noise) and extrinsic noise are identical,
all of the results of the previous section can be carried over
here with the proviso that one replaces the average burst
amplitude $b$ with $b \rightarrow \sigma^2/2\gamma \equiv b_w$ and
$\kappa_b \rightarrow \kappa_e = 2 \gamma \kappa_d /\sigma^2
\equiv \kappa_d/b_w$.

\item  We can look for the regions of bimodality in the
    $(K,\kappa_d)$-plane, for a fixed value of $b_{w }$. We
    have the implicit equation for $x_{\pm}$
    \begin{equation*}
    x^{2n} - (K-1)\left [ n - \dfrac{K+1}{K-1}\right ]x^n -
nb_{w}(K-1)x^{n-1} +K=0
    \label{eq:xpm-implicit-extrinsic}
    \end{equation*}
and the corresponding values of $\kappa_d$ are given by
    \begin{equation*}
    \kappa_{d \pm} = \left( x_{\mp} +b_{w} \right )
    \left(
    \dfrac{K+x_{\mp}^n}{1+x_{\mp}^n}
    \right).
    \label{eq:kappapm-extrinsic}
    \end{equation*}
Then the bimodality region in the $(K,\kappa_d)$-plane with noise in the degradation rate is the same as the
bimodality region for bursting in the $(K,b \kappa_b)$-plane.
    \een
\end{rem}

We have also the following result.

\begin{thm}\cite[Theorem
2]{pichorrudnicki00}. The unique stationary density of Equation
\ref{eq:1d-stoch-decay-fp} is given by Equation
\ref{eq:ss-soln-integ-decay}. Further $u(t,x)$ is asymptotically
stable.
\end{thm}

\subsection{The deterministic limit}

Here again we can recover the deterministic behavior from a limit in the extrinsic fluctuations dynamics. In
this case, however, the frequency and the amplitude of the perturbation are already scaled. Then the limit
$\sigma \to 0$
gives the same result as in the deterministic case.

\section{Discussion and conclusions}\label{sec:disc}


In trying to understand experimentally observed distributions of intracellular
components from a modeling perspective, the norm in computational and systems biology is often to
use algorithms developed initially by \cite{gillespie77} to solve
the chemical master equation for specific situations.  See
\cite{lipniacki} for a typical example. However these
investigations demand long computer runs, are computationally
expensive, and further offer little insight into the possible
diversity of behaviours that different gene regulatory networks
are capable of.

There have been notable exceptions in which the problem has been treated from an analytical point of view, c.f.
\cite{kepler01}, \cite{friedman06}, \cite{rudnicki07}, and \cite{swain08b}.  The
advantage of an analytic development is that one can  determine
how different elements of the dynamics shape temporal and steady
state results for the densities $u(t,x)$ and $u_*(x)$
respectively.

Here we have extended this analytic treatment to simple situations in which there is either bursting transcription  and/or translation (building on and expanding the original work of \citep{friedman06}), or fluctuations in degradation rates, as an alternative to the \citet{gillespie77} algorithm approach.
The advantage of the
analytic approach that we have taken is that it is possible, in
some circumstances, to give precise conditions on the statistical
stability of various dynamics.  Even when analytic solutions are
not available for the partial integro-differential equations governing the
density evolution, the numerical solution of these equations may
be computationally more tractable than using the
\cite{gillespie77} approach.

One of the more surprising results of the work reported here is that the stationary densities in the presence of bursting noise derived in Section \ref{sec:dynamics-single} are analytically indistinguishable from those in the presence of degradation noise studied in Section \ref{sec:dynamics-degrad}.  We had expected that there would be clear differences that would offer some guidance for the interpretation of experimental data to determine whether one or the other was of predominant importance.  Of course, the next obvious step is to examine the problem in the presence of both noise sources simultaneously.  However, the derivation of the evolution equation in this case, as has been pointed out \citep{hierro09}, is not straightforward and we will report on our results in a separate communication.

In terms of the issue of when bistability, or a unimodal versus bimodal stationary density is to be expected, we have pointed out the analogy between the unimodal and bistable behaviours in the deterministic system and the existence of bimodal stationary densities in the stochastic systems.  Our  analysis makes clear the critical role of the dimensionless parameters $n$, $\kappa$ (be it $\kappa_d$,
$\kappa_b$, or $\kappa_e$), $b$ (either $b$ or $b_w$), and the fractional leakage $K^{-1}$.  The relations
between these defining the various possible behaviours are subtle, and we have given these in the relevant
sections of our analysis.

The appearance of both unimodal and bimodal distributions of molecular  constituents as well as what we have
termed Bifurcation Type 1 and Bifurcation Type 2 have been extensively discussed in the applied mathematics
literature (c.f. \citet{horsthemke84}, \cite{ebeling89} and others) and the bare foundations of a stochastic
bifurcation theory have been laid down by \cite{arnold98}.  Significantly, these are also well documented in the
experimental literature as has been shown by \citet{Gardner2000}, \citet{acar05}, \citet{friedman06},
\citet{hawkins2006}, \citet{zach07}, \citet{mariani2010}, and \citet{song2010} for both prokaryotes and
eukaryotes.  If the biochemical details of a particular system are sufficiently well characterized from a
quantitative point of view so that relevant parameters can be estimated, it may be possible to discriminate
between whether these behaviours are due to the presence of bursting transcription/translation or extrinsic
noise.

\section*{Acknowledgements}

This work was supported by the Natural Sciences and Engineering
Research Council (NSERC, Canada), the Mathematics of Information
Technology and Complex Systems (MITACS, Canada), the Alexander von
Humboldt Stiftung, the State Committee for Scientific Research
(Poland) and the Ecole Normale Superieure Lyon (ENS Lyon, France). The work was
carried out at McGill University, Silesian University, the
University of Bremen, and the Oxford Centre for Industrial and
Applied Mathematics (OCIAM), University of Oxford.

\bibliography{zpf2}
\end{document}